\title{Rise or sink: spherical intruder in density frequency dependent static granular fluid}
\author{
Sparisoma Viridi\\
Nuclear Physics and Biophysics Research Division\\
Institut Teknologi Bandung, Bandung 40132, Indonesia\\
\and
Seramika Ari Wahjoedi\\
Theoretical High Energy Physics Reasearch Division\\
Institut Teknologi Bandung, Bandung 40132, Indonesia\\
\and
Suparno Satira\\
Theoretical High Energy Physics Reasearch Division\\
Institut Teknologi Bandung, Bandung 40132, Indonesia\\
\and
Freddy P. Zen\\
Theoretical High Energy Physics Reasearch Division\\
Institut Teknologi Bandung, Bandung 40132, Indonesia\\
}
\date{\today}

\documentclass[12pt]{article}
\usepackage{graphicx}
\usepackage{amssymb}

\begin{document}
\maketitle

\begin{abstract}
A simple model for intruder in vibrating granular bed is constructed by introducing a term that governs frequency dependent granular bed density. Varying the vibrating frequency will drive the buoyant force acting on the intruder by the granular bed. Swapped regions of rising and sinking intruder compared to other reported result have bee found.
\end{abstract}

\section{Introduction}

A phenomenon known as Brazil nut effect (BNE) is introduced descriptively by a work via molecular dynamics (MD) simulation for one and several intruders \cite{Rosato_1987}. The density difference between intruders and granular bed plays important role in size separation of the grains \cite{Moebius_2001}, which determines whether a larger intruder could sink instead of rising \cite{Shinbrot_1998}. Vibrating amplitude, for binary mixture, could drive the state of the mixture into BNE or its reverse (reverse Brazil nut effect, RBNE) \cite{Breu_2003, Ciamarra_2003}. It is interesting to consider that granular bed acts as fluid when the vibrating amplitude and frequency are not zero, but when these parameters are zero, it should be a solid. A simple model that treats the granular bed as fluid is presented with modification to the granular fluid density, which is previously independent from vibrating amplitude and frequency \cite{Viridi_2011}. This model is simpler than that is already reported \cite{Alam_2006}.

\section{Theory}

A spherical intruder with radius $R$ and density $\rho$ is placed in a granular fluid with density $\rho_g$. As it moves updward or downward in the granular fluid it will be under influence of earth gravitational force $F_G$

\begin{equation}
\label{eq01}
F_G = - \frac43 \pi \rho g  R^3,
\end{equation}

\noindent
buoyant force $F_B$

\begin{equation}
\label{eq02}
F_B = \frac43 \pi \rho_g g  R^3,
\end{equation}

\noindent
and viscous drag force $F_D$

\begin{equation}
\label{eq03}
F_D = - 6 \pi \eta R \frac{dy}{dt},
\end{equation}

\noindent
with upward $y$-direction is taken to be positive. Constants $g$ and $\eta$ represent earth gravitational acceleration and granular fluid viscosity, respectively. There are no additional complex forces, such as thermal buoyancy force, compresive force, and dynamic tensile force \cite{Alam_2006}.

Granular fluid density is defined as

\begin{equation}
\label{eq04}
\rho_g = (\rho + \rho_0 \Gamma) \exp{(-\Gamma)},
\end{equation}

\noindent
where $\rho_0$ is density of one particle of granular bed and $\Gamma$ is dimensionless acceleration

\begin{equation}
\label{eq05}
\Gamma = \frac{4 \pi^2 f^2 A}{g},
\end{equation}

\noindent
with $A$ is vibration amplitude and $f$ is vibration frequency, which is common used in indicating the vibration influence \cite{Breu_2003, Ciamarra_2003, Alam_2006}.

Using Newton's second law of motion with Equation (\ref{eq01})-(\ref{eq05}) will give a differential equation

\begin{equation}
\label{eq06}
\frac{d^2y}{dt^2} + c_1 \frac{dy}{dt} + c_2 = 0,
\end{equation}

\noindent
with

\begin{eqnarray}
\label{eq07}
c_1 = \frac{6 \pi \eta R}{m}, \\
\label{eq08}
c_2 = \frac{4 \pi [\rho - (\rho + \rho_0 \Gamma) \exp{(-\Gamma)}] R^3}{3 m}.
\end{eqnarray}

Equation (\ref{eq06}) has solution

\begin{equation}
\label{eq09}
y(t) = \frac{1}{c_1}\left(\frac{c_2}{c_1} - v_0 \right) \exp{(-c_1 t)} + \left(\frac{c_2}{c_1}\right) t + \left[y_0 + \frac{1}{c_1} \left(v_0 - \frac{c_2}{c_1}\right)\right],
\end{equation}

\noindent
where $v_0$ and $y_0$ are initial velocity and position, respectively. Since the intruder always has no initial condition or $v_0 = 0$, then Equation (\ref{eq09}) will be reduced to

\begin{equation}
\label{eq10}
y(t) = \left(\frac{c_2}{c_1^2}\right)\exp{(-c_1 t)} + \left(\frac{c_2}{c_1}\right) t + \left(y_0 - \frac{c_2}{c_1^2}\right).
\end{equation}

\section{Results and discussion}

Because of physical properties of intruder and granular bed Equation (\ref{eq07}) restricts that $c_1 > 0$. For $c_2$ it can be positive or negative according to Equation (\ref{eq08}) since the intruder can have larger or smaller density value than density of one particle of granular bed.

From Equation (\ref{eq10}) velocity of the intruder can be found, which is

\begin{equation}
\label{eq11}
v(t) = \left(\frac{c_2}{c_1}\right)[1 - \exp{(-c_1 t)}],
\end{equation}

\noindent
which tells us that there is a terminal velocity that can have positive and negative value. Positive and negative value means a BNE and a RBNE. respectively. Using Equation (\ref{eq07}) and (\ref{eq08}), Equation (\ref{eq11}) can be rewritten in such form that shows the influence of density difference between intruder and one particle of granular bed, which is

\begin{equation}
\label{eq12}
v(t) \propto [\rho - (\rho + \rho_0 \Gamma) \exp{(-\Gamma)}],
\end{equation}

\noindent
which shows that whether the intruder will rise or sink is dependent explicitely on $\Gamma$.

\begin{figure}[h]
\centering
\begin{tabular}{cc}
\includegraphics[width=6.5cm]{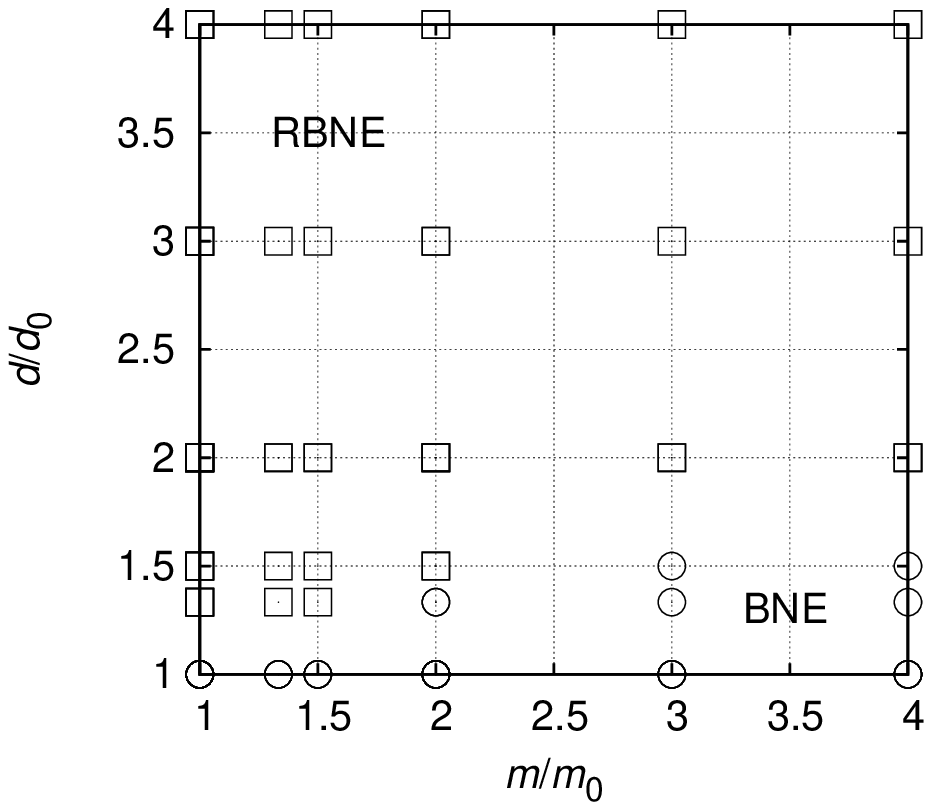} &
\includegraphics[width=6.5cm]{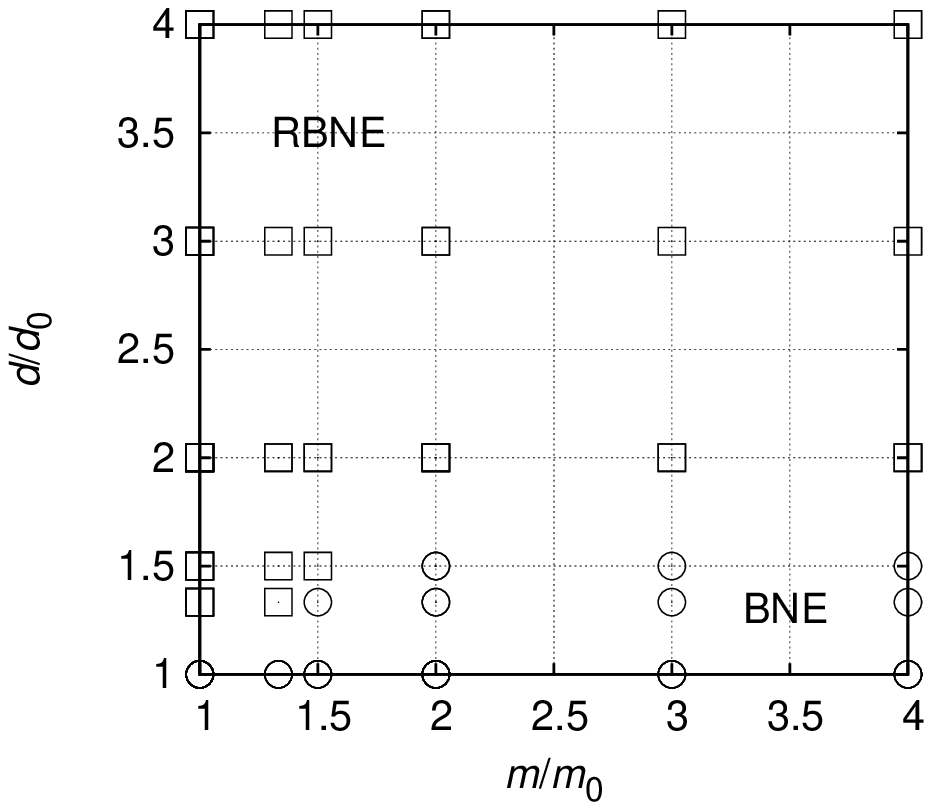} \\
(a) & (b) \\
&\\
\includegraphics[width=6.5cm]{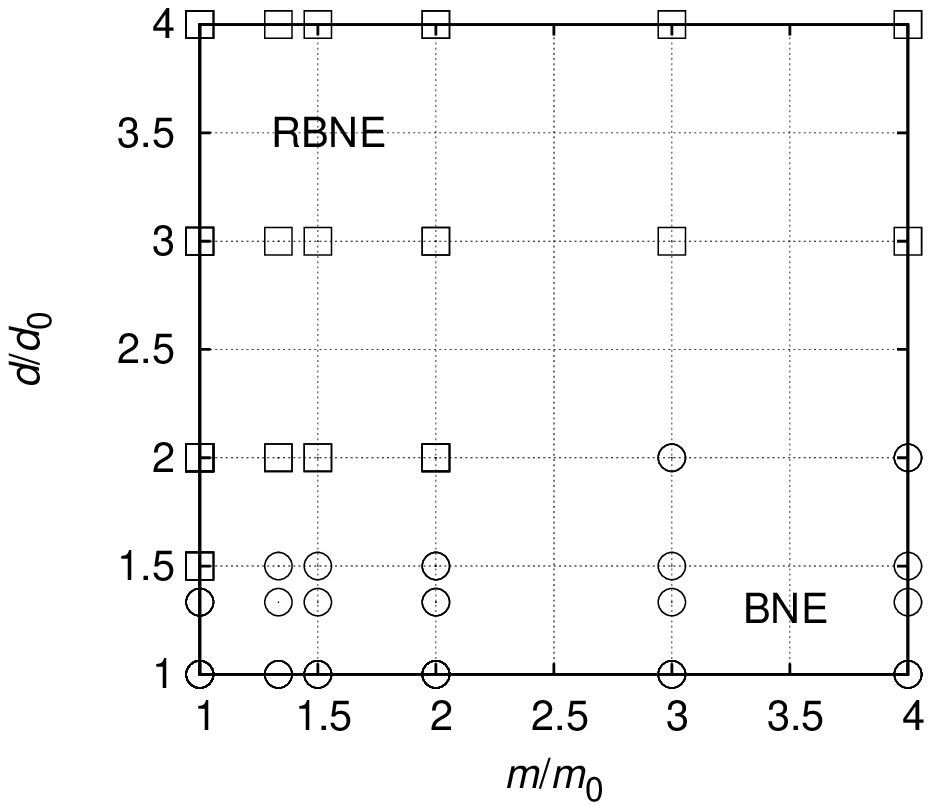} &
\includegraphics[width=6.5cm]{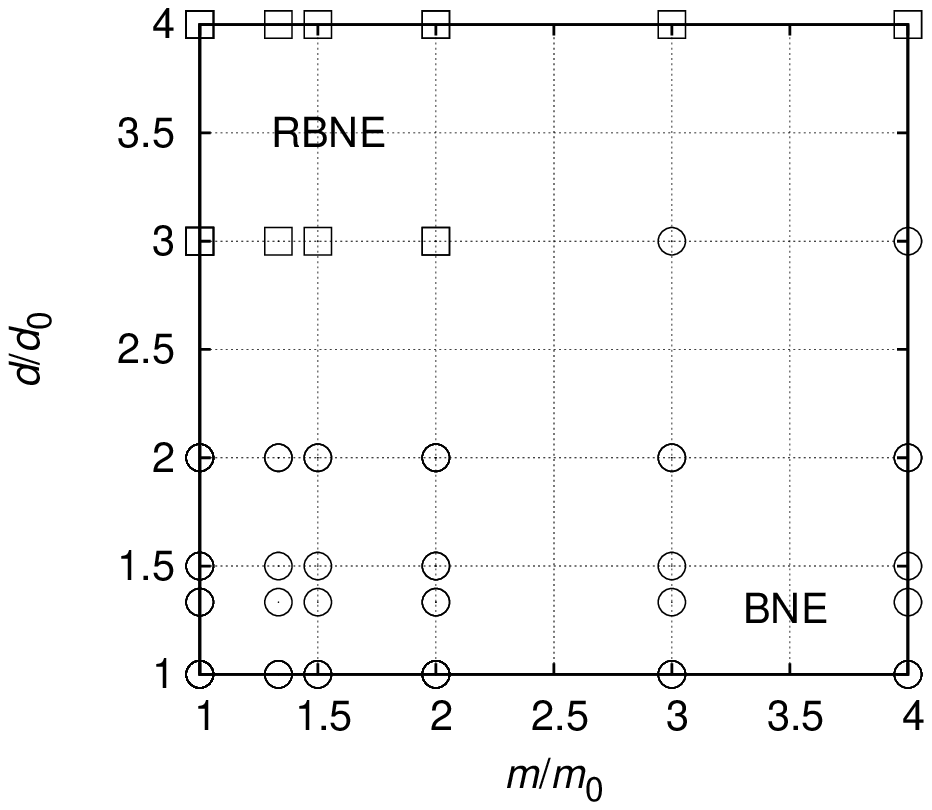} \\
(c) & (d) \\
\end{tabular}
\caption{\label{fg01} Room parameter of $d/d_0$ againts $m/m_0$ for $\Gamma$: (a) 0.5, (b) 1, (c) 2, and (d) 4. BNE and RBNE are indicated by $\circ$ and $\Box$, respectively.}
\end{figure}

Figure \ref{fg01} shows the room parameter of $d/d_0$ againts $m/m_0$ where the regions for BNE and RBNE are swapped as reported in \cite{Hong_2001}, but the dependence of crossover line to $\Gamma$ is similar to what is found by other \cite{Ciamarra_2003}. As $\Gamma$ increases the gradient of transition line in room $d/d_0$ againts $m/m_0$ also increases. For $\Gamma$ about 8 in the range used in Figure \ref{fg01} gives no RBNE states (not shown).

Rise time for the intruder can be found by setting initial and final value of $y$ in the Equation (\ref{eq10}). In this case $y(0) = y_0$ and $y(\tau) = y_f$ will be used to obtain the rise time $\tau$.

\begin{eqnarray}
\nonumber
y_f = \left(\frac{c_2}{c_1^2}\right)\exp{(-c_1 \tau)} + \left(\frac{c_2}{c_1}\right) \tau + \left(y_0 - \frac{c_2}{c_1^2}\right) \\
\nonumber
\Rightarrow y_f - y_0 + \frac{c_2}{c_1^2} = \left(\frac{c_2}{c_1^2}\right)\exp{(-c_1 \tau)} + \left(\frac{c_2}{c_1}\right) \tau \\
\label{eq13}
\Rightarrow 1 + \left(\frac{c_1^2}{c_2}\right)(y_f - y_0) = c_1 \tau + \exp{(-c_1 \tau)}.
\end{eqnarray}

Considered that there is $\tau_0$ where $\tau >> \tau_0$ will reduce Equation (\ref{eq13}) into

\begin{equation}
\label{eq14}
\tau \approx \frac{1}{c_1} + \left(\frac{c_1}{c_2}\right)(y_f - y_0),
\end{equation}

which can be written in

\begin{equation}
\label{eq15}
\tau_{+} \approx \frac{1}{c_1} + \frac{c_3}{[\rho - (\rho + \rho_0 \Gamma) \exp{(-\Gamma)}]},
\end{equation}

\noindent
with

\begin{equation}
\label{eq16}
c_3 = \frac{3 m c_1 (y_f - y_0)}{4 \pi R^3}.
\end{equation}

\noindent
And for $\tau << \tau_0$ will reduce Equation (\ref{eq13}) into

\begin{equation}
\label{eq17}
\tau_{-} \approx -\frac{1}{c_1} \ln{\left\{1 + \frac{c_1 c_3}{[\rho - (\rho + \rho_0 \Gamma) \exp{(-\Gamma)}]}\right\}}.
\end{equation}

\begin{figure}[h]
\centering
\includegraphics[width=10cm]{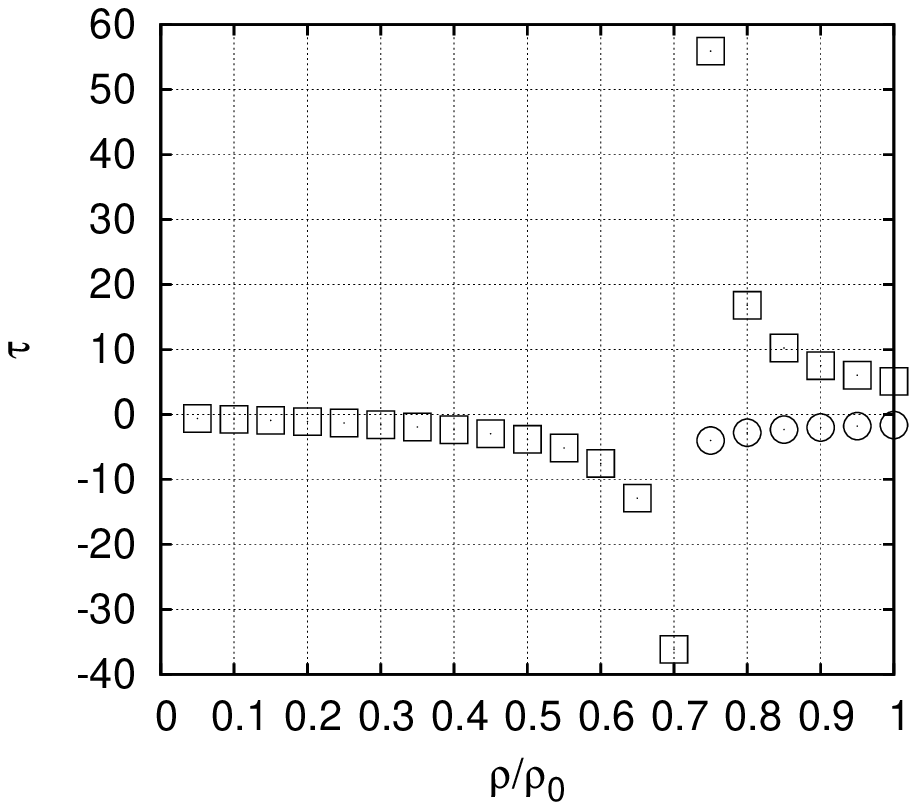}
\caption{\label{fg02} Rise time for $\tau_{-}$ ($\circ$) and $\tau_{+}$ ($\Box$).}
\end{figure}

Figure \ref{fg02} shows unexplained results that there are negative rise time for both calculations using Equation (\ref{eq15}) and (\ref{eq17}). There are two things to do, solve Equation (\ref{eq13}) directly without approximation or redefine Equation (\ref{eq04}) for the granular bed density.

\section{Conclusion}

A model treating granular bed as static fluid has already done. Room parameters for BNE and RBNE shows swapped regions as reported by other. Rise time has also negative values. Further investigation is needed.

\bigskip\bigskip\noindent
{\bf \Large Acknowledgements}\\ \\
Authors would like to thank to Alumni Association Research Grant in year 2010 for financial support to this work.

\bibliographystyle{unsrt}
\bibliography{manuscript}

\end{document}